\documentclass[a4paper]{spie}  %>>> use this instead for A4 paper
\usepackage{amsmath,amsfonts,amssymb}
\usepackage{graphicx}
\usepackage[colorlinks=true, allcolors=blue]{hyperref}

\def\spose#1{\hbox to 0pt{#1\hss}}
\def\simlt{\mathrel{\spose{\lower 3pt\hbox{$\mathchar"218$}}
     \raise 2.0pt\hbox{$\mathchar"13C$}}}
\def\simgt{\mathrel{\spose{\lower 3pt\hbox{$\mathchar"218$}}
     \raise 2.0pt\hbox{$\mathchar"13E$}}}

\def\arcsec{\nobreak{$''$}}

\title{ERIS: revitalising an adaptive optics instrument for the VLT}

\author[a]{R.~Davies}
\author[b]{S.~Esposito}
\author[c]{H.-M.~Schmid}
\author[d]{W.~Taylor}
\author[b]{G.~Agapito}
\author[a]{A.~Agudo~Berbel}
\author[e]{A.~Baruffolo}
\author[b]{V.~Biliotti}
\author[f]{B.~Biller}
\author[d]{M.~Black}
\author[c]{A.~Boehle}
\author[b]{B.~Briguglio}
\author[a]{A.~Buron}
\author[b]{L.~Carbonaro}
\author[a]{A.~Cortes}
\author[b]{G.~Cresci}
\author[a]{M.~Deysenroth}
\author[g]{A.~Di~Cianno}
\author[g]{G.~Di~Rico}
\author[i]{D.~Doelman}
\author[g]{M.~Dolci}
\author[h]{R.~Dorn}
\author[a]{F.~Eisenhauer}
\author[e]{D.~Fantinel}
\author[b]{D.~Ferruzzi}
\author[a]{H.~Feuchtgruber}
\author[a]{N.~F\"orster~Schreiber}
\author[d]{X.~Gao}
\author[a]{H.~Gemperlein}
\author[a]{R.~Genzel}
\author[h]{E.~George}
\author[a]{S.~Gillessen}
\author[b]{C.~Giordano}
\author[c]{A.~Glauser}
\author[h]{A.~Glindemann}
\author[b]{P.~Grani}
\author[a]{M.~Hartl}
\author[h]{J.~Heijmans}
\author[d]{D.~Henry}
\author[a]{H.~Huber}
\author[h]{M.~Kasper}
\author[i]{C.~Keller}
\author[i]{M.~Kenworthy}
\author[c]{J.~K\"uhn}
\author[h]{H.~Kuntschner}
\author[d]{J.~Lightfoot}
\author[d]{D.~Lunney}
\author[d]{M.~MacIntosh}
\author[b]{F.~Mannucci}
\author[c]{S.~March}
\author[h]{M.~Neeser}
\author[c]{P.~Patapis}
\author[d]{D.~Pearson}
\author[a]{M.~Plattner}
\author[b]{A.~Puglisi}
\author[c]{S.~Quanz}
\author[a]{C.~Rau}
\author[b]{A.~Riccardi}
\author[e]{B.~Salasnich}
\author[a]{J.~Schubert}
\author[i]{F.~Snik}
\author[a]{E.~Sturm}
\author[g]{A.~Valentini}
\author[d]{C.~Waring}
\author[a]{E.~Wiezorrek}
\author[b]{M.~Xompero}

\affil[a]{Max Planck Institute for extraterrestrial Physics, 85748 Garching, Germany}
\affil[b]{INAF -- Osservatorio Astrofisico di Arcetri, 50125, Firenze, Italy}
\affil[c]{Institute for Particle Physics and Astrophysics, ETH Zurich, 8093 Zurich, Switzerland}
\affil[d]{UK Astronomy Technology Centre, Edinburgh, EH9 3HJ, UK}
\affil[e]{INAF -- Osservatorio Astronomico di Padova, 35122, Padova, Italy}
\affil[f]{Institute for Astronomy, University of Edinbrugh, Edinburgh, EH9 3HJ, UK}
\affil[g]{INAF -- Osservatorio Astronomico d'Abruzzo, 64100, Teramo, Italy}
\affil[h]{European Southern Observatory, 85748, Garching, Germany}
\affil[i]{Leiden Observatory, University of Leiden, 2300 RA Leiden, The Netherlands}

\authorinfo{Further author information: send correspondence to R. Davies, email davies@mpe.mpg.de}

\begin{document} 
\maketitle

\begin{abstract}
ERIS is an instrument that will both extend and enhance the fundamental diffraction limited imaging and spectroscopy capability for the VLT. It will replace two instruments that are now being maintained beyond their operational lifetimes, combine their functionality on a single focus, provide a new wavefront sensing module that makes use of the facility Adaptive Optics System, and considerably improve their performance. The instrument will be competitive with respect to JWST in several regimes, and has outstanding potential for studies of the Galactic Center, exoplanets, and high redshift galaxies. ERIS had its final design review in 2017, and is expected to be on sky in 2020. This contribution describes the instrument concept, outlines its expected performance, and highlights where it will most excel.
\end{abstract}

% Include a list of keywords after the abstract 

\section{INTRODUCTION}
\label{sec:intro}  % \label{} allows reference to this section

For one and a half decades, the fundamental near-infrared adaptive optics capability for the VLT has been provided by SINFONI and NACO.
SINFONI was installed in 2004 as the combination of the SPIFFI\cite{eis03} integral field spectrometer and MACAO\cite{bon04} adaptive optics system.
SPIFFI itself had been previously operated in 2003 as a seeing limited guest instrument. 
NACO was installed in 2001 as the combination of the CONICA\cite{len03} imager and spectrograph and NAOS\cite{rou03} adaptive optics system.
These two instruments are now working far beyond their 10-year design lifetimes, and so significant investment is needed to maintain the diffraction limit capabilities of the telescope during the coming decade and into the 2030s. 
Furthermore, with the launch of JWST scheduled for March 2021 and the first light of the ELTs planned for the mid-2020s, a significant enhancement of the performance previously offered by SINFONI and NACO is mandatory.

\begin{figure}
\begin{center}
\includegraphics[width=14cm]{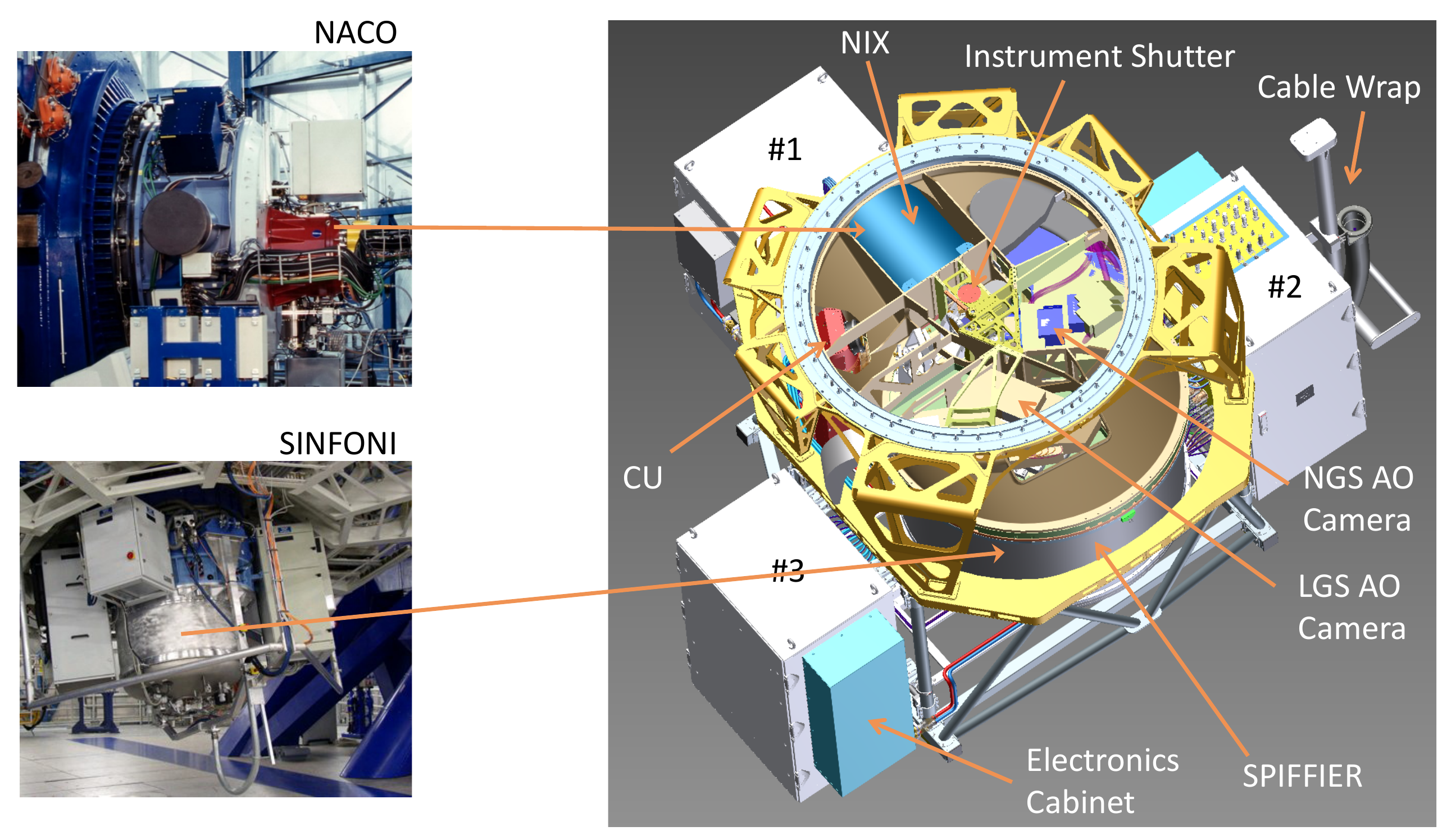}
\end{center}
\caption{ERIS (right) will replace and enhance the capabilities that were previously offered by NACO (left, top) and SINFONI (left, bottom). It will be mounted at the Cassegrain focus of UT4. It comprises a central structure which contains the AO system (which will make use of the lasers, adaptive secondary mirror, and real-time computer of the AOF), a high contrast imager NIX, an integral field spectrograph SPIFFIER, a calibration unit, and various electronics.}
\label{fig:intro}
\end{figure}

ERIS, the Enhanced Resolution Imaging Spectrograph, will achieve this. 
As indicated in Fig.~\ref{fig:intro}, it is a Cassegrain instrument that will combine a replacement for NAOS (CONICA), a full refurbishment and upgrade to SINFONI (SPIFFI), and a new adaptive optics module that makes use of the Adaptive Optics Facility\cite{ars17} (AOF) on UT4. 
A preliminary upgrade of SPIFFI was carried out in January 2016, and the completed ERIS instrument will be on-sky in 2020.

Following a report by ESO at the end of 2010, the ERIS project was originally conceived to fill a window of opportunity to develop an instrument with imaging and spectroscopic capabilities, that would exploit the AOF.
The associated Phase A study was successfully completed in 2012.
After this, the project was temporarily put on hold, during which time it was re-organised, enabling Phase B to begin at the end of 2014.
The consortium now comprises the following members: Max-Planck-Institut f\"ur extraterrestrische Physik (the PI institute, responsible for system engineering, central structure, and SPIFFIER); three INAF institutes, Osservatorio Astrofisico di Arcetri (adaptive optics), Osservatorio Astronomico d'Abruzzo (calibration unit), and Osservatorio Astronomico di Padova (control software); UK Astronomy Technology Centre (NIX); Eidgen\"ossische Technische Hochschule Z\"urich (NIX wheels with their filters and masks); Leiden Observatory (high contrast imaging); and the European Southern Observatory (detectors and handling tool).
Following a mandate from ESO's Scientific Technical Committee that, given the development landscape at the time, a first light for the instrument after 2020 would be scientifically less compelling, the project pushed through its initial phases quickly. 
The Preliminary Design Review was held in February 2016, and the Final Design Review in May 2017.
The project is now in its manufacturing and assembly phase.
One important `fixed point' in the schedule is the decommissioning of SINFONI, which is planned for mid 2019 following the most critical observations of the S2 star around its current closest approach to the massive black hole in the Galactic Center.
It is planned that the imaging camera NIX, as well as the adaptive optics system in the central structure will be integrated and tested by that time. Following a rapid refurbishment of SPIFFI, and its re-integration as SPIFFIER into ERIS, the complete instrument will be shipped back to the observatory at the beginning of 2020, ready for commissioning.

\section{Primary Science Drivers and Capabilities}
\label{sec:science}

ERIS is a workhorse instrument optimised for diffraction limited observations of individual targets.
Because it offers general purpose capabilities to that end, it can be used for a wide variety of different science cases. 
Here, we highlight three priority science topics, which exploit the instrument’s capabilities and illustrate the regimes in which ERIS is competitive with JWST. 

\subsection{Galaxy Evolution at High Redshift}
\label{sec:highz}

\begin{figure}
\begin{center}
\includegraphics[width=16cm]{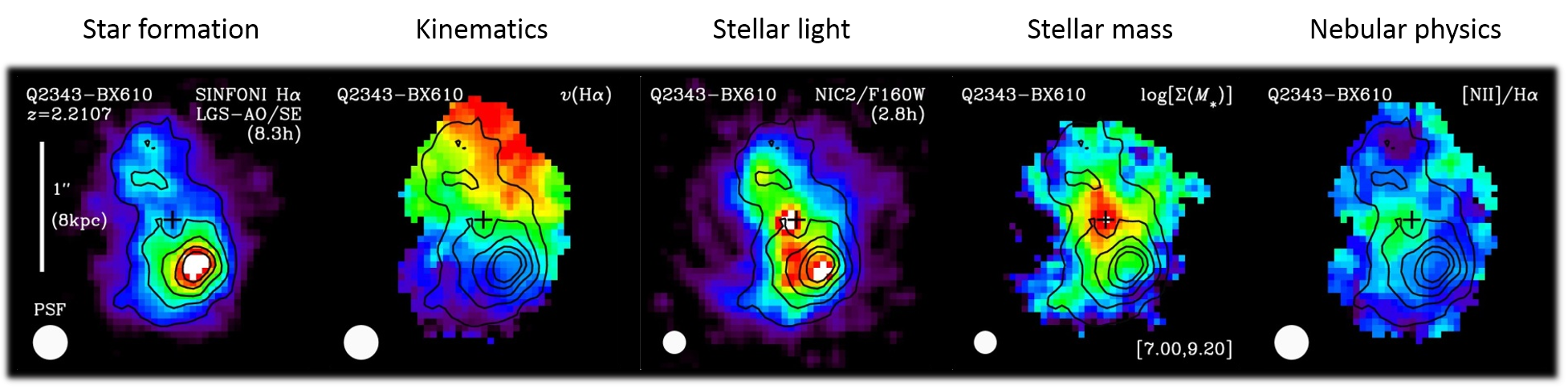}
\end{center}
\caption{Images of Q2343-BX610 at $z = 2.21$ showing structures from different tracers at $\sim 1$\,kpc spatial resolution\cite{for11}. The star formation and kinematics traced by H$\alpha$, and nebular physics traced by the [NII]/H$\alpha$ ratio are from SINFONI. The analysis is enhanced by combining these with distributions of the stellar light and mass from HST.}
\label{fig:highz}
\end{figure}

The focus of this science case is on the signatures of the physical processes driving mass assembly and structural transformations of galaxies at redshifts $z \sim 1$--3, which includes look-back times around the peak of the cosmic star formation rate density as well as the subsequent shutdown of star formation. 
These processes include the growth of bulges, inflows in disks, star formation in and between clumps, as well as feedback and quenching from star formation and AGN. 
And their signatures can be seen most clearly by their imprint on galaxy kinematics, which occurs on scales of $<1$\,kpc and 10--40\,km/s, as illustrated in Fig.~\ref{fig:highz}.

Currently, the number of galaxies that have been observed with adaptive optics integral field spectroscopy in this redshift range is limited\cite{for18}, and covers a mixed bag of objects, typically bright and/or having high specific star formation rate and/or strongly lensed. 
During the last few years, multi-object instruments like KMOS\cite{sha13} have begun to provide a census of thousands of such galaxies\cite{wis15}, but at seeing limited scales of $\sim5$\,kpc. 
ERIS will reveal the detailed structures of galaxies in these samples.
To do so, high spatial resolution of 0.1\arcsec\ or better is required to spatially resolve these processes from each other, as well as to study compact galaxies. 
High spectral resolution of $R \sim 4000$--8000 is required to trace the kinematics, especially due to the decline over cosmic time of the characteristic intrinsic velocity dispersion of disks (setting their vertical thickness), and to probe efficiently between the bright sky OH lines. Wavelength coverage in the J, H, and K bands is required to observe key
emission lines at the redshifts of interest; 
and high sensitivity is required to do so in low mass galaxies and to characterise higher order moments of emission lines.

ERIS will be an important facility for such work, because no current or planned space-based instrumentation will provide the necessary spectral resolution for kinematic studies of galaxies. 
Indeed, JWST has a single IFU with only $R < 3000$; and its focus will be on multi-object spectroscopy of faint galaxies, measuring multi-line diagnostics for a census of galaxy populations up to the highest redshifts. 
Exploiting its AO performance, sensitivity in the JHK bands, and high resolution grating, ERIS will have the potential to surpass JWST in terms of probing the physical mechanisms of galaxy evolution and star formation shutdown.

\subsection{Direct Imaging of Exoplanets}
\label{sec:exoplanet}

\begin{figure}
\begin{center}
\includegraphics[width=12cm]{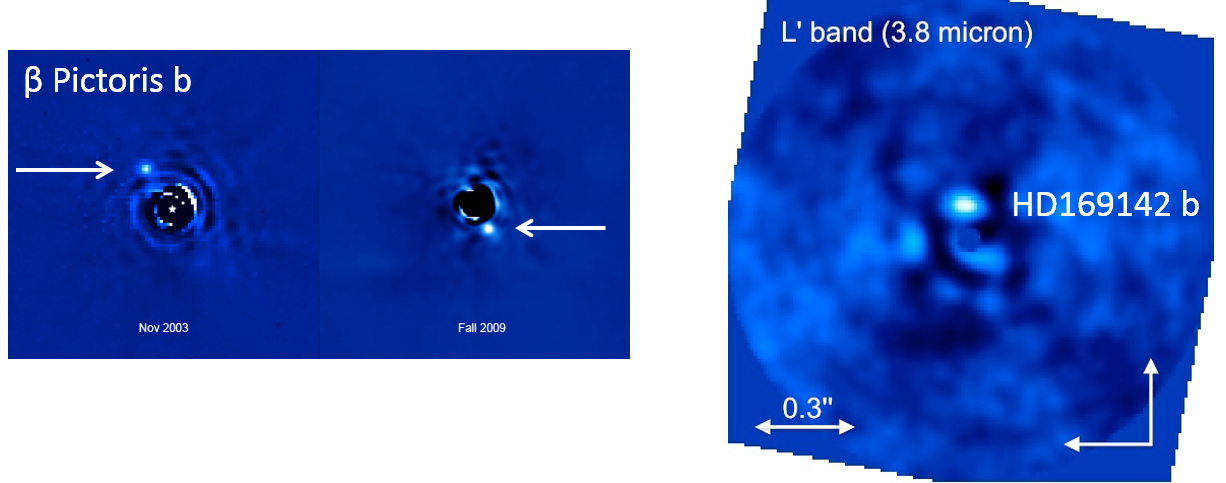}
\end{center}
\caption{Key L-band detections of exoplanets made by NACO. Left: $\beta$\,Pictoris\,b is a 4--11\,M$_{Jup}$ planet orbiting at a distance of 9\,AU from the star in a plane close to that of the debris disk at larger scales\cite{lag10,bon13}. Right: HD169142\,b is a protoplanet that has not been observed at shorter wavelengths, located in the transition disk around the star\cite{reg14,bil14}. SAMs and coronagraphs, in combination with Angular Diffrential Imaging (ADI), significantly ease the detection of such objects.}
\label{fig:lband}
\end{figure}

The 3--5\,$\mu$m wavelength regime, covering the L- and M-bands, plays a central role in the discovery and characterisation of exoplanets, as the examples in Fig.~\ref{fig:lband} show. 
Indeed, before SPHERE\cite{beu08} and GPI\cite{mac14}, most detections of exoplanets were in the L-band.
And this remains true for the colder -- i.e. older or lower mass -- planets\cite{vig15}. 
It is also the case at the other end of the age range, where longer wavelengths are highly competitive for detecting young proto-planets that are still highly obscured. 
The warm (few 100\,K) circumplanetary material in which they are embedded provides a large emitting area, making coronagraphy and sparse aperture masking in the L- and M-bands efficient tools to characterise them\cite{qua13,qua15}.

\begin{figure}
\begin{center}
\includegraphics[width=16cm]{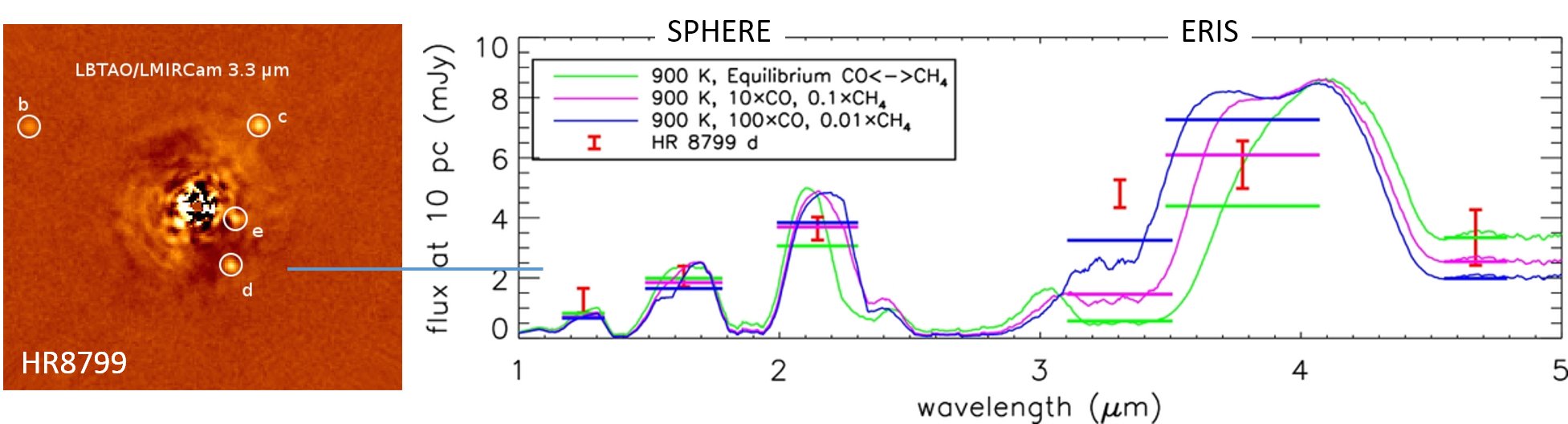}
\end{center}
\caption{Illustration, using HR8799\,d as an example, of how ERIS can contribute to resolving questions about atmosphere chemistry in exoplanets\cite{ske12}. Left: L-band image of the HR8799 system indicating the four known planets. Right: JHK photometry from SPHERE cannot always distinguish equilibrium and non-equilibrium models, while possible LM measurements from ERIS may provide more decisive diagnostic power.}
\label{fig:hr8799}
\end{figure}

In addition to detecting very young and old planets, ERIS provides a unique probe of the atmospheres of gas giant planets, particularly in the context of disequilibrium chemistry and clouds.
Combining SPHERE and GPI detections in the JHK bands with follow-up observations by ERIS in the LM bands will vastly improve the efficacy of the scientific analysis by providing a longer baseline for fitting atmosphere models. 
As illustrated in Fig.~\ref{fig:hr8799}, for the case of two planets in the HR8799 system (comprising 4 gas giant planets of 7--10\,M$_{Jup}$) standard atmosphere models cannot explain the observed photometry, particularly in the L-band. 
Such work is in its infancy, and the sensitivity and 3--5\,$\mu$m capabilities of ERIS open the potential for exciting major advances.

JWST excels in terms of background limited sensitivity due to the low background in space, but it is still speckle limited at small angular separations. 
The PSF is very complex and even with a coronagraph there will be significant residual light on small radial scales.
On the other hand, ERIS has coronagraphs that are optimised for high throughput and a small inner working angle.
The two facilities will therefore be complementary, probing different regimes of planetary characteristics.

\subsection{Galactic Center}
\label{sec:gc}

The Galactic Center is considered a priority science case for ESO.
And the vast improvements over the last decade in spatial resolution (to 50\,mas, equivalent to 0.002\,pc) and sensitivity (by 3--5\,mag) made possible by AO on the VLT, have revealed surprises and challenge theories of star formation and gas inflow in galaxy nuclei\cite{gen10}. 

Some of the discoveries that have come about via astrometric imaging and integral field spectroscopy -– capabilities that lie at the core of ERIS -– include the differing spatial distributions of late type stars, early stars, and Wolf-Rayet stars, as well as the various 3D orbital structures they trace. Frequent and regular monitoring of the orbits of the stars closest to Sgr\,A* at unprecedented precision for more than two decades has provided exquisite constraints on both the mass of and distance to the supermassive black hole\cite{gil17}.
The discovery of a gas cloud falling towards Sgr A*, and the tracking of its subsequent tidal disruption, generated considerable interest in the community, as well as numerous theories about what it is and where it has come from. 
While visible at L-band, it has not been detected at K-band, putting strong constraints on the dust temperature. 
It is visible most spectacularly in recombination lines such as Br$\gamma$.
Interestingly, the change in velocity shear over time observed in the ionised gas can be well matched by a simple set of test particles moving in the black hole’s potential, suggesting that the ionised gas is not bound together. 
The instrumental requirements to observe the clouds -– L-band imaging and H/K band integral field spectroscopy -– are among those for which ERIS is optimised. 
And that such an event was seen so soon after the capability to observe it was developed suggests that ERIS has a good chance of observing future unexpected and serendipitous events that will teach us more about the evolution of galaxy nuclei and accretion processes onto their supermassive black holes.

In the Galactic Center, the key science topics (dynamics of the various stellar populations, the radiative behaviour of Sgr\,A*, flares and gas streamers, and continued monitoring of fainter and closer stars around Sgr\,A*) are those where ERIS will be highly competitive with respect to JWST. 
For stellar orbits, the most critical criterion is resolution, for which larger mirror size and shorter wavelengths (to H-band, beyond which extinction is prohibitive) are both important. 
Since sensitivity is limited by crowding, the lower background of JWST yields little advantage. 
Furthermore, the regular monitoring required over the long term is well suited to ground-based observatories but would strain the scheduling of a space-based telescope with a limited fuel supply.

\section{System Overview}
\label{sec:overview}

ERIS comprises several sub-systems which are labelled in Fig.~\ref{fig:intro} and described in this section:
\begin{itemize}
\item The central structure, to which the other sub-systems are mounted.
\item The AO system\cite{ric16,ric18}, which is a distributed system, with wavefront sensors mounted in the central structure, and making use of the Adaptive Optics Facility on UT4.
\item NIX\cite{pea16}, the 1--5\,$\mu$m imager (which effectively replaces NACO).
\item SPIFFIER, the integral field spectrograph (a major upgrade of SPIFFI).
\item The calibration unit\cite{dol16,dol18}, which allows internal calibration and registration of wavefront sensors, NIX (up to 2.5\,$\mu$m; 3--5\,$\mu$m calibrations must be done on sky), and SPIFFIER.
\item The instrument control system, comprising the control network and the instrument software.
\item The handling tool, which enables ERIS to be removed from the telescope for maintenance.
\end{itemize}

\subsection{Central Structure}
\label{sec:central}

\begin{figure}
\begin{center}
\includegraphics[width=15cm]{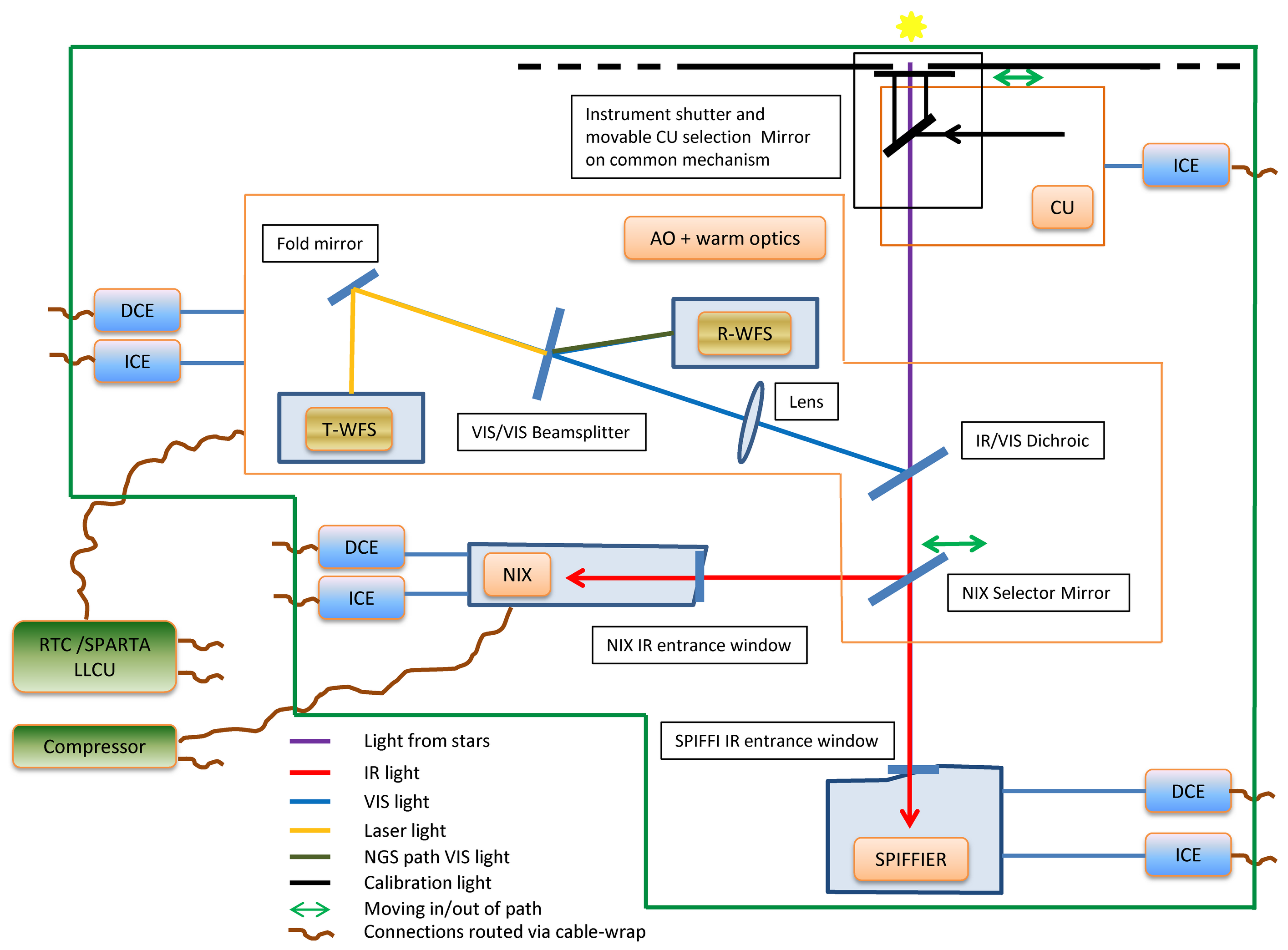}
\end{center}
\caption{Sketch showing the concept for the optics in the central structure: how the light from the calibration unit is folded in when the instrument shutter is closed; that the optical path goes directly to SPIFFIER passing only the IR/VIS dichroic which splits light off to the wavefront sensors; how a fold mirror can be moved in to reflect the light instead to NIX. Electronics connections are also shown.}
\label{fig:censtruct}
\end{figure}

The central structure is the module that mounts ERIS to the Cassegrain adaptor/rotator and contains the warm optics. 
In order to achieve the high throughput and low thermal background required for ERIS, there is no optical relay. 
Instead, as illustrated in Fig.~\ref{fig:censtruct}, the light from the telescope is either transmitted directly into SPIFFIER or reflected into NIX, passing only the infrared/visible dichroic that splits off the optical light to the wavefront sensors.
To enable this requires the back focal length of the telescope to be extended from 0.25\,m to 0.50\,m, implying some associated modifications to the guider arm and camera.

The topmost element in the central structure is the shutter unit which is closed either when ERIS is not being used, or during internal calibrations. 
As such, it has a fold mirror (calibration unit selector mirror) mounted to its underside which reflects light from the calibration unit into the beam path.
This fold mirror has to be at the start of the optical path, since the calibration unit serves SPIFFIER, NIX, and the AO system.

Immediately below this unit is the infrared/visible (or instrument) dichroic, which is fixed in position.
Due to the volume restriction for the optical train to the wavefront sensors, the dichroic has to be tilted at 45$^\circ$. 
The backside of the optic is shaped to correct the resulting astigmatism. 
But the tilt also leads to the cameras seeing additional thermal background due to the small fraction of infrared light reflected by the dichroic.
To offset this and achieve the specification on low thermal background, a turret extension from the SPIFFIER cryostat has been made with an additional window so that there is a cold plate directly behind the dichroic.

Below this is a movable fold mirror (the NIX selector mirror).
This is switched out when the light is fed to SPIFFIER, and moved in when the light should be fed to NIX. 

These three elements fill the 0.3\,m height envelope available within the central structure. 
The rest of the limited volume within the central structure is occupied by the wavefront sensors, described in Sec.~\ref{sec:ao}.

\subsection{Adaptive Optics System}
\label{sec:ao}

\begin{figure}
\begin{center}
\includegraphics[width=12cm]{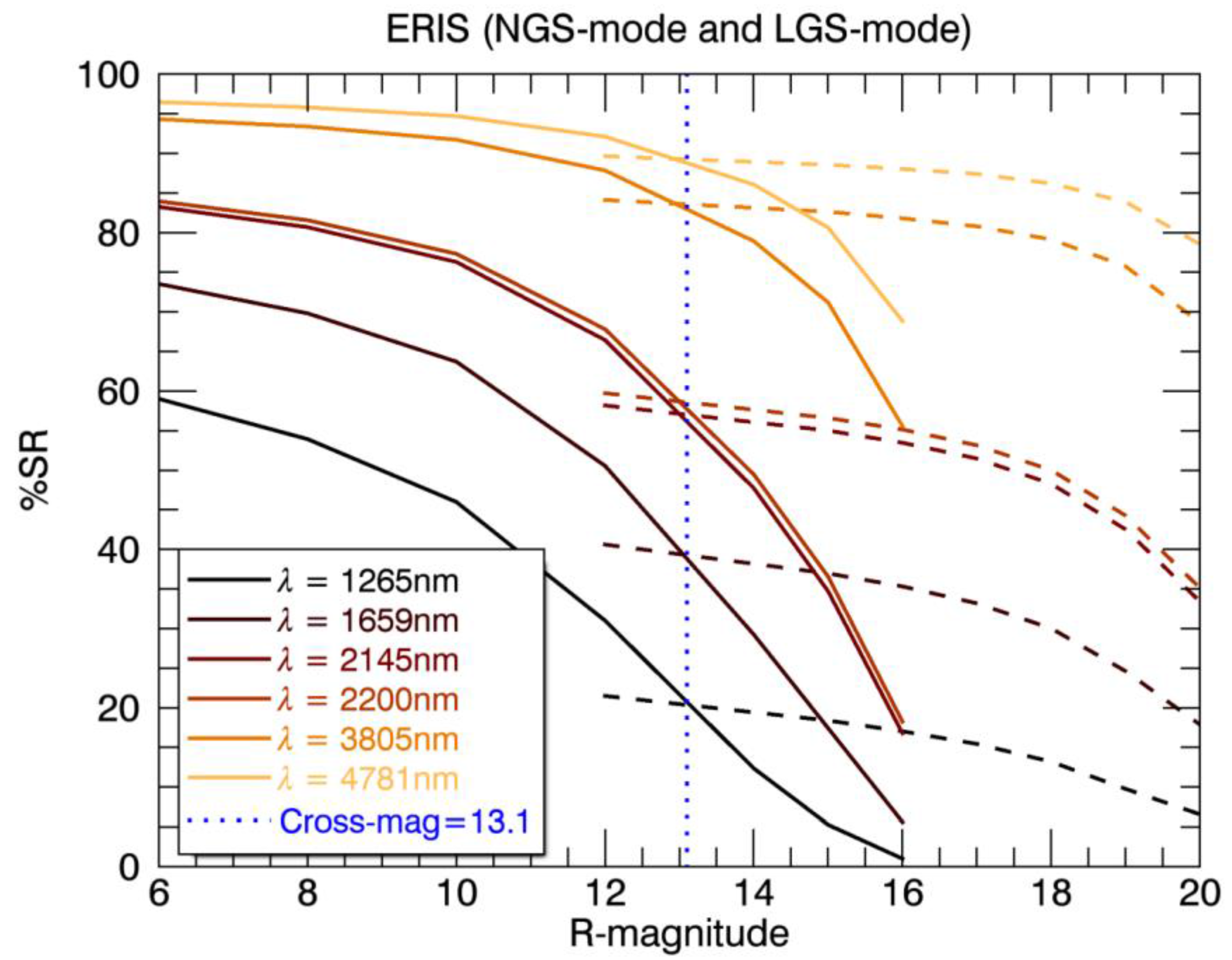}
\end{center}
\caption{AO performance of ERIS in J, H, K, L, and M bands with natural (solid) and laser (dashed) guide stars. 
The latter provides better performance for NGS fainter than 13.1\,mag in R-band, assuming no anisoplanatism.
These calculations are far above what was achieved with NACO or SINFONI.
In L-band, Strehls suitable for high contrast imaging will be achievable in LGS-AO mode using tip-tilt stars as faint as 18\,mag. 
Even in J-band, very respectable performance will be achievable with faint stars, vastly increasing the potential of this band for high resolution studies across a wide range of topics.}
\label{fig:aoperf}
\end{figure}

In ERIS, the AO system\cite{ric16,ric18} comprises the warm optics and the wavefront sensors, as illustrated in Fig.~\ref{fig:censtruct}.
The optical train to the WFS shows that, after the infrared/visible dichroic, there is an additional visible/visible dichroic which transmits the narrow band 589\,nm sodium LGS wavelength and reflects the rest of the optical light. 
The transmitted light is folded into a high order WFS that is fixed on-axis and is optimised to sense the wavefront from the LGS. 
The reflected light goes to a NGS WFS that can work both in high order for NGS-AO, and in low order for LGS-AO. 
It is movable so it can pick up the reference star off-axis, and also follow the reference source either when non-sidereal targets are observed or during off-axis guiding in pupil tracking mode.

The AO system will make use of key elements of the Adaptive Optics Facility\cite{ars17} (and hence ERIS has to be mounted on UT4). 
In particular, any one of the four lasers can be used to
generate a LGS (when required), real-time computations make use of the SPARTA\cite{fed06,sua12} infrastructure (ESO's standard platform for adaptive optics real time applications), and the wavefront correction is applied to the deformable secondary mirror.
The performance of the AO system, shown in Fig.~\ref{fig:aoperf} for all the relevant bands, has been calculated using these elements. 
It indicates that with either NGS or LGS, one can achieve a Strehl ratio in L-band exceeding 80\% with a reference star as faint as 18\,mag. 
This has a huge potential for coronagraphic observations of sub-stellar and planetary companions around stars. 
Similarly, the K-band performance is significantly higher than in either NACO or SINFONI.
Indeed with bright stars, the expected ERIS performance is comparable to the specification for JWST of 80\% Strehl at 2\,$\mu$m; and with stars as faint as 17\,mag it will not be far below this level. 
Crucially, ERIS will also provide valuable AO performance in J-band. 
Coupled with the significantly enhanced J-band performance of SPIFFIER, this will open up new scientific opportunities that were previously inaccessible.

\subsection{NIX Imager}
\label{sec:nix}

\begin{figure}
\begin{center}
\includegraphics[width=15cm]{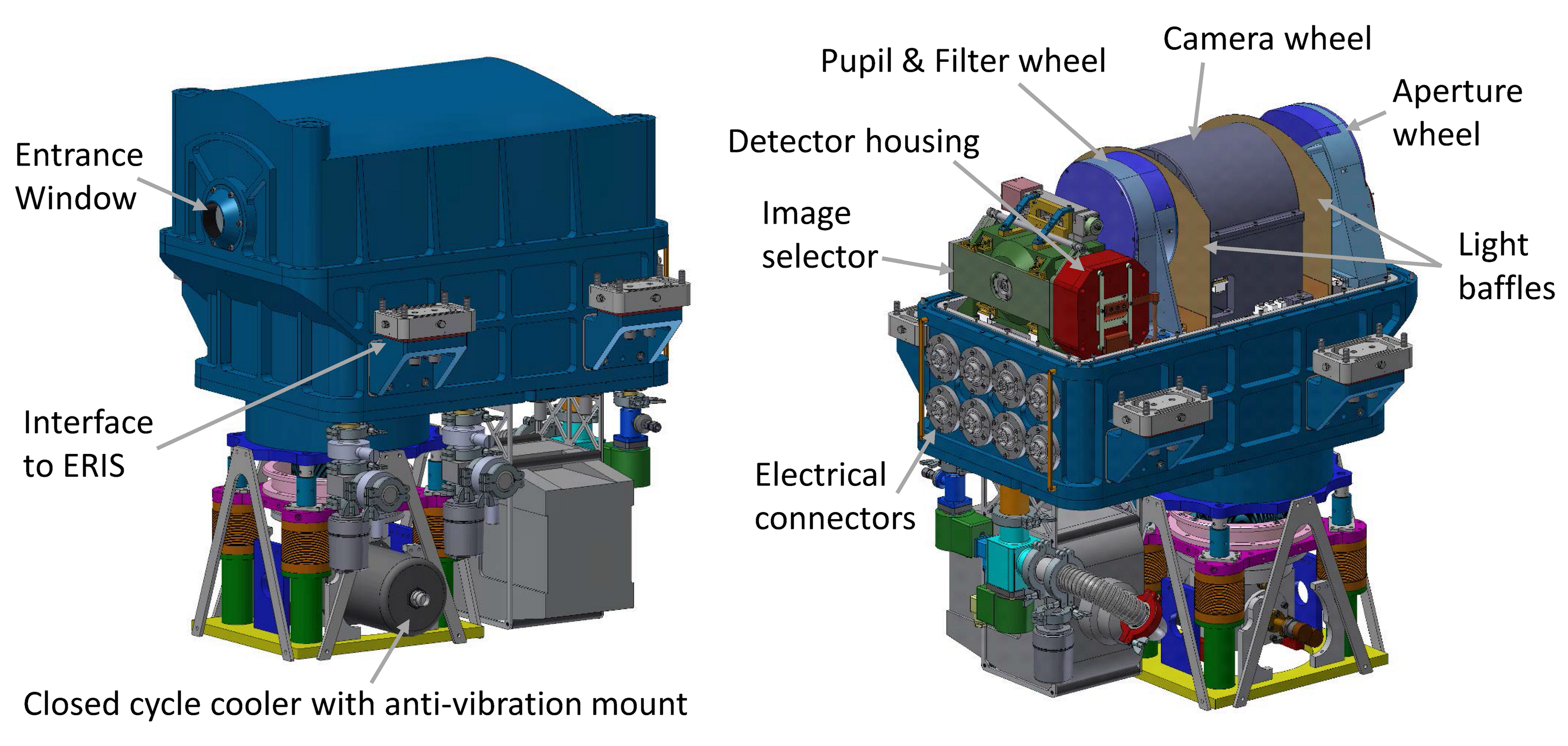}
\end{center}
\caption{Overview of the NIX imager. Views are from the front and back, with the cryostat cover in place and removed.}
\label{fig:nix}
\end{figure}

The NIX imager\cite{pea16}, shown in Fig.~\ref{fig:nix}, is a replacement for NACO. 
It provides diffraction limited imaging from J- to M-band over a field of up to 30\arcsec\ with a 13\,mas pixel scale. 
In addition, for J- to K-bands, a larger pixel scale of 27\,mas is available, providing almost a full 1\,arcmin field. 
An important capability is high contrast imaging, and this is enabled by pupil and focal plane coronagraphs, as well as sparse aperture masks. 
In addition, there is a simple long-slit spectroscopic mode covering the full L-band.
This provides essentially diffraction limited performance along the full 12\arcsec\ length of the slit, with a resolution of $R \sim 450$ for an 86\,mas slit width.

The optical design is, in principle, relatively simple. 
After the cryostat window is an aperture wheel which can switch different optical elements (imaging masks, focal plane coronagraphs, slit) into the beam\cite{gla18}.
Following this is the camera lens barrel which can be switched between three options according to the required pixel scale and wavelength range. 
A second mechanism holds the pupil wheel (masks, stops, grisms, neutral density filters) and filter wheel (contains only filters)\cite{gla18}. 
The light is then directed to the detector unit via fold mirrors. 
The varying numbers of optical elements along the path leads to the necessity for two focus positions of the detector that are separated by 2.5\,mm.
The detector unit therefore has an adjustable focus stage. 
Apart from the new mounting for the H2RG, the unit is otherwise similar to those used for other VLT instruments, and an NGC will be used to read out the
detector.

A key part of NIX is the cooling system, which is a closed cycle cooler.
Despite the obvious concerns about vibration, the rationale for this choice includes the experience within the consortium with such devices, and the successful vibrationally isolated implementation on other instruments such as KMOS. 
In addition, a single system can fulfil all the temperature requirements; while the alternative of liquid nitrogen would not be able to cool the detector unit to the required temperature and so an additional cooling system (e.g. a pulse tube cooler) would anyway be needed and lead to the same
concerns. 
The ability to tune the damping of the isolation unit so as to remove specific key frequencies, and the additional use of filtering in the adaptive optics control algorithms will ensure that vibration can be controlled to a sufficient level.

\begin{figure}
\begin{center}
\includegraphics[width=14cm]{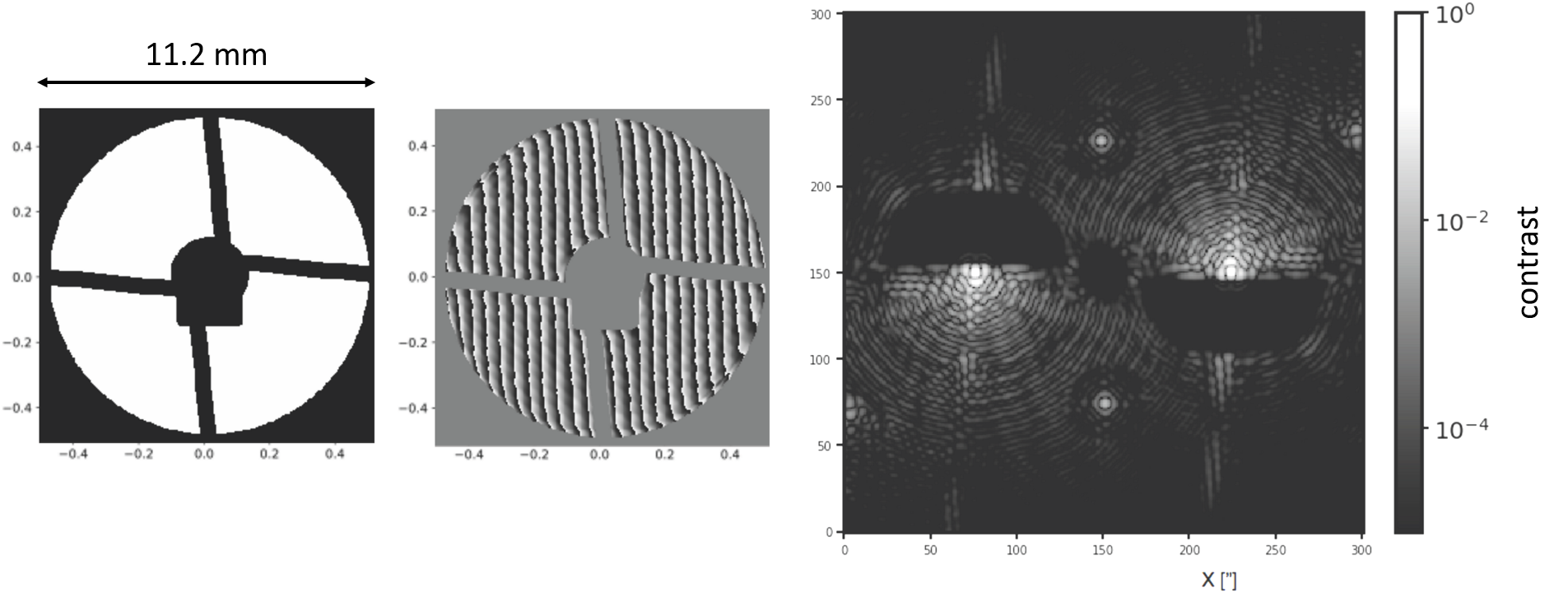}
\end{center}
\caption{Left: Mask used for the grating vector apodised phase plate (gvAPP)\cite{ken18} located in the NIX cold pupil, marking the spider arms and deformable secondary mirror baffle, which is oversized in order to block thermal emission from M3 in its folded position. Centre: the phase distribution used for the phase plate. Right: the resulting image expected. Three PSFs are generated with the gvAPP. The undeviated PSF contains about 2\% of the total flux in a non-coronagraphic PSF and acts as a photometric and astrometric reference for the other two coronagraphic PSFs (this leakage component is not included above). These two PSFs combine to provide full 360 degree suppression around the target star. An additional two holographic PSFs provide simultaneous measurement of the focus of the gvAPP on the science detector.}
\label{fig:gvapp}
\end{figure}

The two primary science requirements for NIX are astrometric imaging and high contrast imaging. To fulfil the former, the camera has been designed to have very low optical distortion. For the latter, a suite of phase masks is included.
In the focal plane, two annular grove phase masks (AGPM or vortex) will be available.
These have high throughput, and are good for extended objects such as circumstellar disks as well as exoplanets. 
On the other hand, precise centering is important, which is achieved using the QACITS algorithm\cite{hub17} which provides the necessary feedback from the science exposures to the control software.
In the pupil plane, a grating vector apodised phase mask (gvAPP) will be available, as shown in Fig.~\ref{fig:gvapp}.
This single device will operate from K to M band.
The mask has been manufactured and is currently undergoing laboratory tests\cite{ken18,boe18}.
These can be custom designed for a variety of purposes, and have been demonstrated on sky\cite{ott14,ott17}.

\subsection{SPIFFIER Integral Field Spectrograph}
\label{sec:spiffier}

SPIFFIER is an upgrade and refurbishment of SPIFFI, which is the IFU camera of SINFONI.
The functionality will remain basically the same, but with significantly improved throughput and image quality. 
Operationally, SPIFFIER will also remain basically the same as SPIFFI. 
The main difference is that the low resolution H+K grating will be replaced by a high resolution grating providing $R \sim 8000$, covering about half a band at a time within the J, H, and K atmospheric windows. 
This will be offered at 3 fixed settings for each band, rather than allowing the central wavelength to be freely adjustable.

\begin{figure}
\begin{center}
\includegraphics[width=11cm]{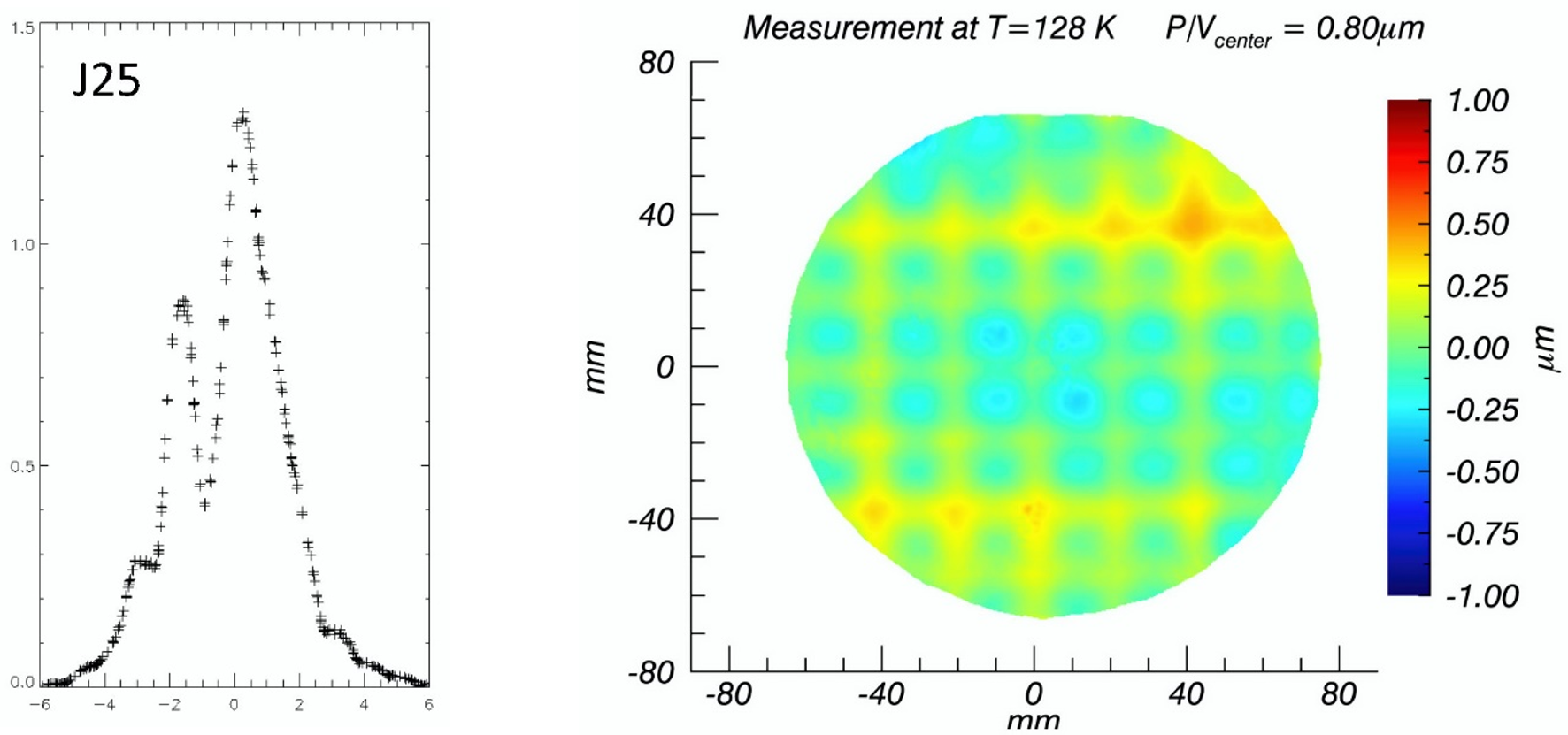}
\end{center}
\caption{Left: example of a particularly poor J-band spectral line profile measured in SINFONI, super-sampled to show the full detail. Right: a major contribution to the complex profile is the light-weighting of the gratings. When they are cooled, this leads to an `eggbox' structure that mimics a grating\cite{geo17}. The gratings will be replaced when SPIFFI is refurbished as SPIFFIER and integrated into ERIS.}
\label{fig:spiffier}
\end{figure}

To achieve the instrument’s full potential in terms of sensitivity and spectral resolution (and to provide a clean spectral line profile), the collimator mirrors were already exchanged at the beginning of 2016\cite{geo16} and, when refurbished for integration into ERIS, all the gratings will also be exchanged.
As Fig.~\ref{fig:spiffier} shows, the rationale for the latter step is that the light-weighting of the current gratings leads to an `eggbox' deformation when cooled, which acts as an an additional grating and leads to a complex spectral profile\cite{geo17}.
With new collimator mirrors and gratings, as well as filters (those installed in 2016 have a throughput exceeding 98\% across the full band\cite{geo16}), the sensitivity is expected to increase by a significant factor.
In addition, the entrance window will be replaced by a smaller one matched to the nominal field (since the sky spider has been removed), and the pre-optics collimator will be replaced to accommodate the new AO optical interface.
The detector will also be replaced.
Based on an initial report, the new detector should have good performance in terms for a number of key paraemters such as quantum efficiency, read noise, cosmetics, and dark current. 
And the first measurements of persistence suggest that it should be improved with respect to the current detector\cite{geo16}.
However no definitive statement on relative performance of the current versus new detectors will be possible before the side-by-side test that is currently in the planning.
Finally, the motors, electronics, and detector will also be replaced, as will the cryostat lid and some internal baffling.

\subsection{Calibration Unit}
\label{sec:cu}

\begin{figure}
\begin{center}
\includegraphics[width=16cm]{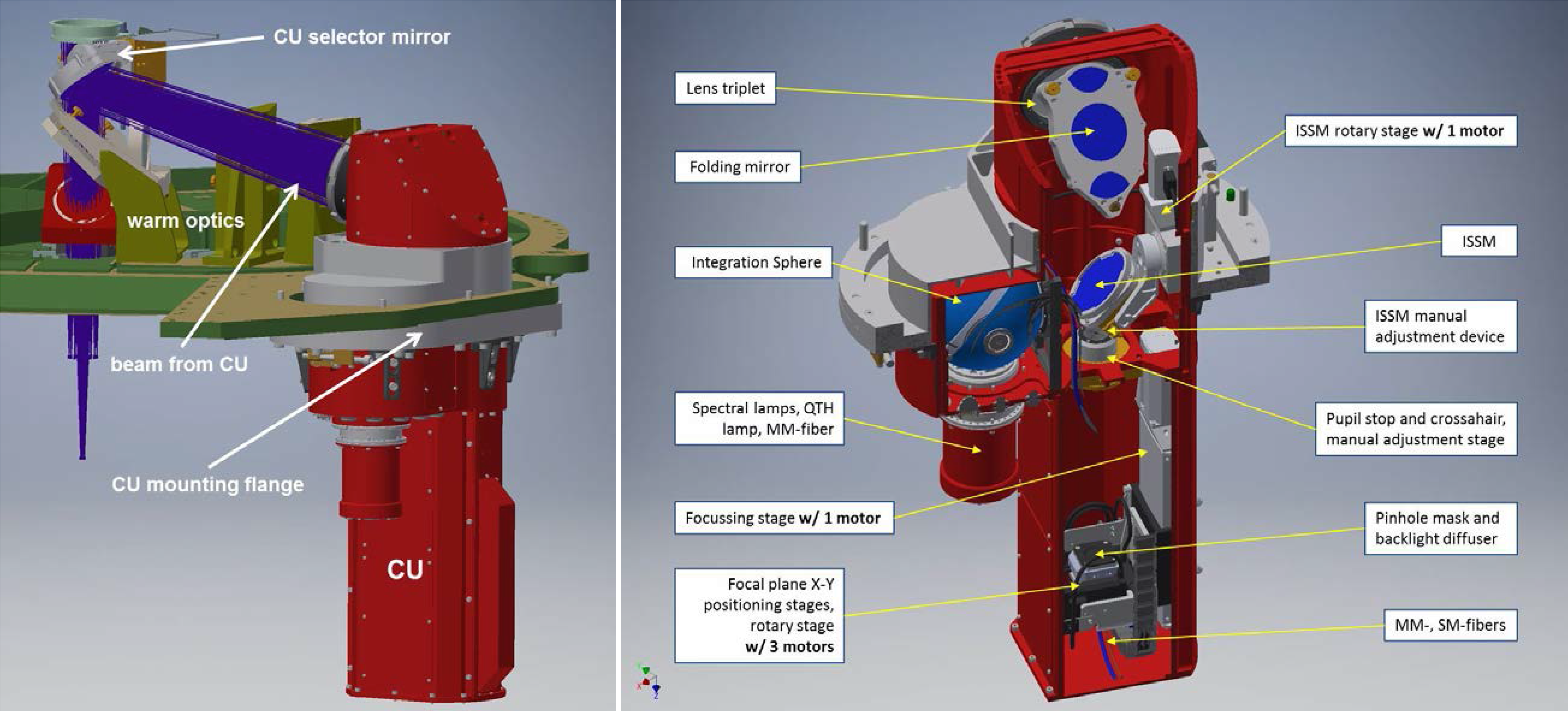}
\end{center}
\caption{Calibration unit main bench. Left: mounted to ERIS. Right: showing the internal opto-mechanics. The light sources are either from the integrating sphere (for uniform illumination) or through the pinhole mask (for point-like or extended sources, fed from the fibre switchyard). In the latter case, one positioning stage can move the source across the field of view, while another can change its focus position to simulate either NGS or LGS. The ISSM (integration sphere selector mirror) switches between the sources. A fold mirror then directs the light to the triplet lens which focusses it onto the telescope focal plane.}
\label{fig:cu}
\end{figure}

The calibration unit\cite{dol16,dol18} will provide internal sources for flatfielding and wavelength calibration from J-band to K-band, as well as optical-to-infrared point sources for AO calibration, including non-common path aberrations to the instrument detector focal planes. 
Calibrations required by NIX at 3--5\,$\mu$m have to be done on-sky. 
This is necessary due to opto-mechanical constraints (available
volume, transmissivity of optics at all required wavelengths), and is considered acceptable from the operational and scientific perspectives.

The calibration unit comprises two main units: the main bench, shown in Fig.~\ref{fig:cu}, which mounts to the central structure as indicated in Fig.~\ref{fig:censtruct}; and the fibre switchyard, located in an electronics cabinet and interfaces to the main bench. 
Neutral density filters at various locations allow the adjustment of intensity levels.
The primary functionality of the calibration unit is to provide:
\begin{itemize}
\item
Halogen lamp and laser diode light source via an integrating sphere, to enable flatfields for NIX in broad and narrow band filters as well as spectroscopic flatfields for SPIFFIER (both in the range 1--2.5\,$\mu$m);
\item
Ar, Ne, Kr, and Xe arclamps covering the same wavelength range to calibrate SPIFFIER dispersion solution;
\item
Sets of 3 illuminated slits for calibrating SPIFFIER slitlet curvature and alignment at 1--2.5\,$\mu$m;
\item
Point-like and extended sources (diffraction limited to 1.5\arcsec\ diameter) at optical (R-band) wavelengths, for AO calibration of both NGS and LGS (including the effects of the finite 80--200\,km distance to the LGS);
\item
Point-like source (with an almost flat spectrum from 0.4--2.4\,$\mu$m) for aligning the instruments and wavefront sensors, measuring differential flexure, and deriving non-common path aberrations between them.
\end{itemize}

\subsection{Instrument Control System}

\begin{figure}
\begin{center}
\includegraphics[width=12cm]{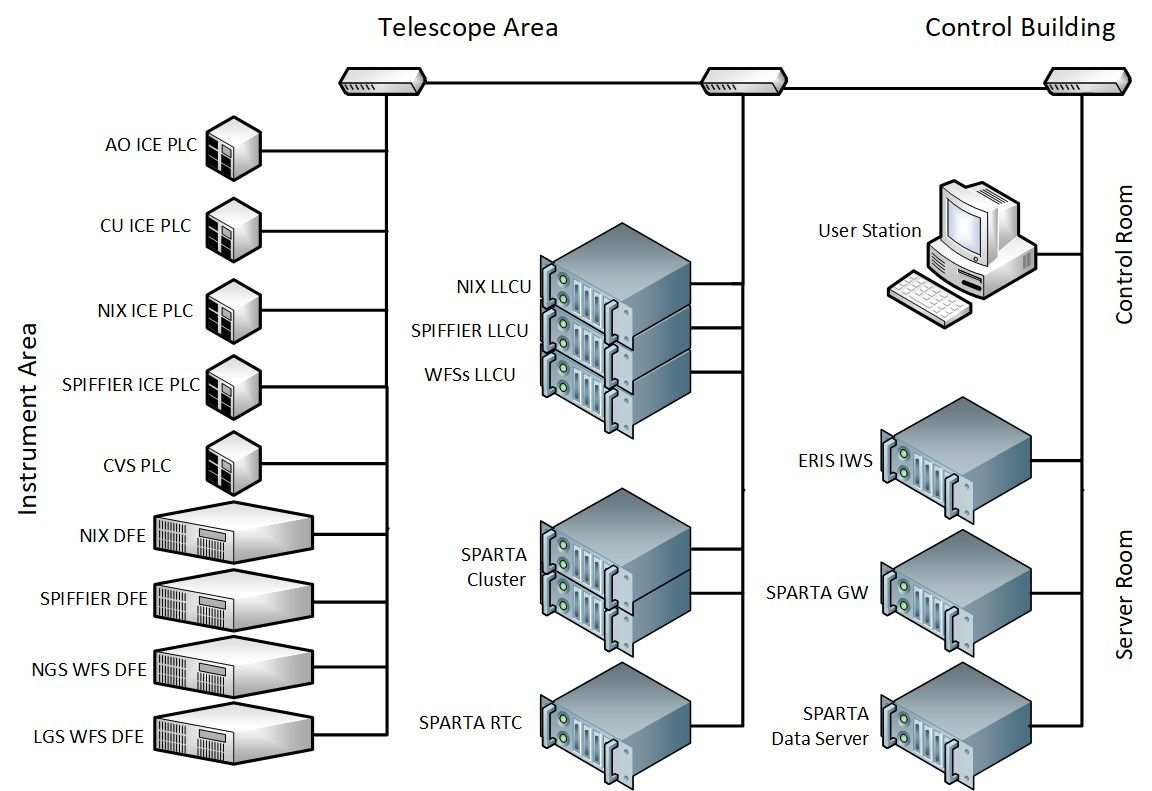}
\end{center}
\caption{Conceptual architecture of the ERIS control network.}
\label{Fig:Ctrl-Net}
\end{figure}

The main responsibility of the Instrument Control System is to put all detectors and functions under remote control, so that observation, calibration and maintenance operations can be executed by means of software. 
Additionally, it must monitor all sensors and manage the cryo-vacuum system.

Fig.~\ref{Fig:Ctrl-Net} shows a conceptual view of the ERIS control network.
Since the instrument will be installed at the VLT, its control system must adhere to the ESO standards\cite{9152E} both in hardware and in software. As such, function control is based on a network of Programmable Logic Controllers (PLCs), while science detectors are controlled by the ESO New General Controller (NGC\cite{NGC}), comprising the Detector Front-end Electronics (DFE) and Linux Local Control Units (LLCUs). 
The control network also includes the machines for Adaptive Optics control, based on ESO's SPARTA platform~\cite{fed06,sua12} and, finally, the Instrument Work-Station (IWS). 
This is where all coordinating, monitoring and user interface software resides; and where all operational software, implemented in the form of templates, runs. 
Companion papers describe the ERIS control system\cite{10703-129} (in particular for the Adaptive Optics and Calibration Unit subsystems) and the design of ERIS Instrument Software\cite{10707-52}.

\subsection{Handling Tool}
\label{sec:handling}

The handling tool will be mainly used for handling the instrument in the integration facility and mounting it to the Cassegrain flange of the telescope, as well as during maintenance, or recoating of the M1 mirror.
Since the ERIS instrument has a stiff support structure for the electronics cabinets, no support of the instrument will be needed during the integration
of the instrument. 
The concept for the handling tool is thus a platform, which is supporting the entire instrument from the bottom. 
By raising the platform, the instrument can be lifted and attached to the Cassegrain flange. 
To allow flexible access to the screws at the Cassegrain flange, the platform will not include any working area for people. 
Instead, a dedicated scaffolding with wheels will be used.
For moving the instrument in the integration facility and place it precisely under the Cassegrain flange, a motorized drive will be included in the system handling tool.

\acknowledgments % equivalent to \section*{ACKNOWLEDGMENTS}       
 
The consortium thanks the staff at the partner institutes, at ESO in Garching, and at ESO Paranal, for their continued support, which has made the project possible. RD thanks M.~Meyer for his enthusiasm for the project, especially during its early phases.

% References
%\bibliography{report} % bibliography data in report.bib
\bibliographystyle{spiebib} % makes bibtex use spiebib.bst

\end{document}